\documentclass{eptcs}
\providecommand{\event}{23rd International Conference on Quantum Physics and Logic (QPL 2026)} 

\usepackage{underscore}
\usepackage[T1]{fontenc}
\usepackage{mathtools}
\usepackage{cleveref}
\usepackage{enumerate}
\usepackage{algorithm}
\usepackage{algpseudocode}
\usepackage{caption}
\usepackage{subcaption}
\usepackage{booktabs}

\usepackage{tikzit}
\include{quantum.tikzdefs}

\tikzstyle{box}=[shape=rectangle, text height=1.5ex, text depth=0.25ex, yshift=0.5mm, fill=white, draw=black, minimum height=5mm, yshift=-0.5mm, minimum width=5mm, font={\small}]
\tikzstyle{gate}=[shape=rectangle, text height=1.5ex, text depth=0.25ex, yshift=0.5mm, fill=white, draw=black, minimum height=5mm, yshift=-0.5mm, minimum width=5mm, font={\small}, tikzit category=circuit]
\tikzstyle{big gate}=[shape=rectangle, text height=1.5ex, text depth=0.25ex, yshift=0.5mm, fill=white, draw=black, minimum height=10mm, yshift=-0.5mm, minimum width=5mm, font={\small}, tikzit category=circuit]
\tikzstyle{Z dot}=[inner sep=0mm, minimum size=2mm, shape=circle, draw=black, fill={rgb,255: red,221; green,255; blue,221}, tikzit category=zx]
\tikzstyle{Z phase dot}=[minimum size=5mm, font={\footnotesize\boldmath}, shape=rectangle, rounded corners=2mm, inner sep=0.2mm, outer sep=-2mm, scale=0.8, tikzit shape=circle, draw=black, fill={rgb,255: red,221; green,255; blue,221}, tikzit draw=blue, tikzit category=zx]
\tikzstyle{X dot}=[Z dot, shape=circle, draw=black, fill={rgb,255: red,255; green,136; blue,136}, tikzit category=zx]
\tikzstyle{X phase dot}=[Z phase dot, tikzit shape=circle, tikzit draw=blue, fill={rgb,255: red,255; green,136; blue,136}, font={\footnotesize\boldmath}, tikzit category=zx]
\tikzstyle{hadamard}=[fill=yellow, draw=black, shape=rectangle, inner sep=0.6mm, minimum height=1.5mm, minimum width=1.5mm, tikzit category=zx]
\tikzstyle{paulibox}=[fill={rgb,255: red,221; green,221; blue,255}, draw=black, shape=rectangle, inner sep=0.6mm, minimum height=5mm, minimum width=5mm, font={\footnotesize}, text height=1.5ex, text depth=0.25ex, tikzit category=zx]
\tikzstyle{vertex}=[inner sep=0mm, minimum size=1mm, shape=circle, draw=black, fill=black, tikzit category=misc]
\tikzstyle{vertex set}=[inner sep=0mm, minimum size=1mm, shape=circle, draw=black, fill=white, font={\footnotesize\boldmath}, tikzit category=misc]
\tikzstyle{small black dot}=[fill=black, draw=black, shape=circle, inner sep=0pt, minimum width=1.2mm, tikzit category=circuit]
\tikzstyle{cnot ctrl}=[fill=black, draw=black, shape=circle, inner sep=0pt, minimum width=1.2mm, tikzit category=circuit]
\tikzstyle{cnot targ}=[fill=white, draw=white, shape=circle, tikzit category=circuit, label={center:$\oplus$}, inner sep=0pt, minimum width=2.1mm, tikzit fill={rgb,255: red,102; green,204; blue,255}, tikzit draw=black]
\tikzstyle{ket}=[fill=white, draw=black, shape=regular polygon, regular polygon sides=3, regular polygon rotate=-30, scale=0.7, inner sep=1pt, tikzit category=circuit, tikzit shape=rectangle, tikzit fill=green]
\tikzstyle{bra}=[fill=white, draw=black, shape=regular polygon, regular polygon sides=3, regular polygon rotate=30, scale=0.7, inner sep=1pt, tikzit category=circuit, tikzit shape=rectangle, tikzit fill=red]
\tikzstyle{scalar}=[shape=rectangle, text height=1.5ex, text depth=0.25ex, yshift=0.5mm, fill=white, draw=black, minimum height=5mm, yshift=-0.5mm, minimum width=5mm, font={\small}]
\tikzstyle{clabel}=[fill=white, draw=none, shape=rectangle, tikzit fill={rgb,255: red,56; green,255; blue,242}, font={\footnotesize}, inner sep=1pt, tikzit category=labels]
\tikzstyle{empty diagram}=[draw={gray!40!white}, dashed, shape=rectangle, minimum width=1cm, minimum height=1cm, tikzit category=misc]

\tikzstyle{simple}=[-]
\tikzstyle{hadamard edge}=[-, dashed, dash pattern=on 2pt off 0.5pt, thick, draw={rgb,255: red,68; green,136; blue,255}]
\tikzstyle{box edge}=[-, dashed, dash pattern=on 2pt off 0.5pt, thick, draw={rgb,255: red,203; green,192; blue,225}]
\tikzstyle{brace edge}=[-, tikzit draw=blue, decorate, decoration={brace,amplitude=1mm,raise=-1mm}]
\tikzstyle{diredge}=[->]
\tikzstyle{double edge}=[-, double, shorten <=-1mm, shorten >=-1mm, double distance=2pt]
\tikzstyle{gray edge}=[-, {gray!60!white}]
\tikzstyle{pointer edge}=[->, very thick, gray]
\tikzstyle{boldedge}=[-, line width=1.6pt, shorten <=-0.17mm, shorten >=-0.17mm]

\title{Classical Clifford+T Sampling without Computing Marginals}
\author{Mark Koch
\institute{Department of Computer Science\\
University of Oxford}
\email{mark.koch@cs.ox.ac.uk}
}
\def\titlerunning{Classical Clifford+T Sampling without Computing Marginals}
\def\authorrunning{M. Koch}

\newcommand{\pvec}[1]{\vec{#1}\mkern2mu\vphantom{#1}}

\usepackage[amsthm,amsmath,thmmarks]{ntheorem}
\theoremstyle{plain}
\newtheorem{lemma}{Lemma}

\theoremstyle{nonumberplain}

\begin{document}
\maketitle

\begin{abstract}
	We introduce a new exact classical sampling algorithm for Clifford+T circuits, called T-by-T sampling.
	For a circuit with T-count $t$, it reduces sampling to evaluating $2t$ amplitudes of circuits with successively increasing T-counts $1,2,...,t$, without the need to compute costly marginal probabilities.
	Crucially, the number of amplitude evaluations is completely independent of the surrounding Clifford structure.
	This makes T-by-T sampling well suited for use together with low-rank stabiliser decompositions, which enable efficient amplitude evaluations as long as the T-count is not too high.
	In that sense, T-by-T sampling is an improvement over the gate-by-gate sampling technique introduced by Bravyi, Gosset, and Liu~\cite{bravyi2022how}, where the number of amplitudes instead scales with the number of non-monomial operations such as Hadamard gates, which can remain large even in circuits with low stabiliser rank.
	Concretely, T-by-T sampling offers significant speedups in regimes where T gates are embedded within large Clifford computations.
	This makes it well suited for applications such as simulating quantum error correction experiments, which typically have lower T complexity but involve extensive Clifford operations and measurements.
	In particular, we show how T-by-T sampling allows us to take samples from recent end-to-end magic state cultivation circuits with escape up to distance 19 in a couple of seconds.
\end{abstract}

\section{Introduction}

It is widely believed that simulating universal quantum circuits on a classical computer has an exponential runtime cost.
Nevertheless, classical simulation of quantum systems continues to play an important role in the field, with applications ranging from the design and validation of quantum algorithms to benchmarking and verification of hardware.
Consequently, there is significant interest in accelerating classical simulation methods and coming up with techniques that work well for specific classes of circuits.
One such example is the class of \emph{Clifford+T circuits} that allow for fault-tolerant compilation while at the same time being approximately universal.
As a result, Clifford+T serves as the standard compilation framework for many fault-tolerant architectures, and the number of T gates, also known as the \emph{T-count}, is widely regarded as a key resource measure.
An effective technique for simulating Clifford+T circuits with relatively low T-counts is the \emph{low-rank stabiliser decomposition} method~\cite{bravyi2016improved, bravyi2019simulation}.
Given a circuit $C$, the idea is to decompose $C\ket{0}$ into a linear combination of stabiliser states.
The number of terms in this decomposition, also known as the \emph{stabiliser rank}, is believed to scale as $2^{\alpha t}$ where $t$ denotes the T-count of $C$ and $\alpha < 1$ is a constant.
The best known strategy achieves $\alpha \approx 0.396$ and was discovered in \cite{qassim2021improvedupperbounds, kissinger2022classical}.
This decomposition then allows us to compute an output probability $P(\vec x) := \left| \bra{\vec x}C\ket{0} \right|^2$ by evaluating the corresponding amplitude for each stabiliser term and summing up the contributions.
The task of computing such probabilities $P(\vec x)$ is also known as \emph{strong simulation}.

In many applications, however, the objective is not to compute individual probabilities, but rather to \emph{sample} from the output distribution $P$.
This task is also known as \emph{weak simulation}.
For example, in the context of quantum error correction (QEC), one typically performs Monte Carlo simulations over many noisy shots to estimate the logical error rate of fault-tolerant protocols or gadgets.
A standard method to reduce weak simulation to strong simulation proceeds via sequential marginalisation.
The idea is to calculate marginal probabilities $P_j(x_1, ..., x_j) = \bra{0}C^\dagger(\ket{\vec x}\bra{\vec x} \otimes I_{n-j}) C\ket{0}$ and then sample the qubit outcomes one by one, conditioning on the previously sampled bits and marginalising over the ones that have not been sampled yet.
However, this approach comes with a significant runtime overhead.
In the case of Clifford+T circuits, computing a marginal $P_j$ corresponds to evaluating an amplitude of a circuit with \emph{twice} the number of T gates.
This leads to an exponential runtime increase from $O(2^{\alpha t})$ to $O(4^{\alpha t})$, making it very costly in practice.

Several approaches have been proposed to mitigate this overhead.
For example, Refs. \cite{bravyi2016improved, bravyi2019simulation} approximate marginal probabilities via a norm estimation technique based on overlaps with carefully chosen random stabiliser states.
\cite{bravyi2019simulation} also proposed a Metropolis-Hastings Markov Chain Monte Carlo method for approximate sampling, however without rigorous error bounds.
To the best of our knowledge, the current state-of-the-art method is the \emph{gate-by-gate} algorithm proposed by Bravyi, Gosset, and Liu~\cite{bravyi2022how}.
The high-level idea is to begin with a trivial sample $\vec x = 0$ and update it by applying the gates of the circuit one by one, using amplitude evaluations to resample affected bits.
Concretely, if $C$ is expressed in the CNOT+Hadamard+T gate set, then the number of required amplitude evaluations will scale linearly with the number of Hadamard gates.
While this avoids the doubling of the T-count inherent in marginal-based sampling, the number of Hadamard gates can still be very large even for circuits with low stabiliser rank.
For example, this situation is typical in QEC circuits, which are dominated by extensive Clifford subroutines, such as syndrome extraction, lattice surgery etc., while non-Clifford resources are primarily concentrated in magic state preparation.
As a consequence, the gate-by-gate method may still require a large number of expensive amplitude evaluations despite a relatively low T-count.

These considerations motivate the development of a sampling method in which the number of amplitude evaluations depends primarily on the T-count, rather than the surrounding Clifford operations.
This work introduces a new exact sampling algorithm for Clifford+T circuits that achieves precisely this goal.
The new method, which is called \emph{T-by-T sampling}, requires only $2t$ amplitude evaluations for a circuit with T-count $t$, completely independent of the Clifford structure of the circuit.
Moreover, most of these amplitude evaluations involve circuits with significantly smaller T-count than $t$.
Concretely, the algorithm proceeds by evaluating amplitude pairs of circuits with increasing T-counts $1, 2, ..., t$, so that only the very last step incurs the full strong simulation cost of $O(2^{\alpha t})$.
As a result, T-by-T sampling provides a substantial practical speed-up for circuits where T gates are embedded within large Clifford computations.
After reviewing the gate-by-gate method in \Cref{sec:gate-by-gate} and introducing T-by-T sampling in \Cref{sec:t-by-t}, this advantage is demonstrated through numerical benchmarks in \Cref{sec:experiments}, including experiments on random circuits as well as magic state cultivation circuits from the recent QEC literature~\cite{gidney2024cultivation}.

\section{Background: Gate-by-Gate Sampling} \label{sec:gate-by-gate}

Before introducing T-by-T sampling, it will be useful to first review Bravyi, Gosset, and Liu's \cite{bravyi2022how} gate-by-gate method in detail.
To this end, let $C = U_m\cdots U_1$ be a quantum circuit (not necessarily Clifford+T) consisting of a sequence of $m$ unitary gates $U_i$.
The goal is to sample from the distribution $P(\vec x) := \left| \bra{\vec x}C\ket{0} \right|^2$.
Let $C_k = U_k\cdots U_1$ be the subcircuit containing only the first $k$ gates.
In particular, we have $C_0 = I$ and $C_m = C$.
Finally, define the output distribution at time step $k$ as $P_k(\vec x) := \left|\bra{\vec x} C_k \ket{0}\right|^2$.
In particular, $P_0(\vec x) = \left|\braket{\vec x}{0}\right|^2 = \delta_{\vec x = 0}$ and $P_m = P$.
The key idea of gate-by-gate sampling is to start with a sample from the trivial distribution $P_0$ (i.e. the all-zero bit string) and updating it one gate at a time until we arrive at a sample that is distributed according to $P_m = P$.
This works because adjacent distributions $P_{k-1}$ and $P_k$ are closely related: they only differ by a single gate $U_k$ that moreover only acts on a small subset $S \subseteq \{1, ..., n\}$ of qubits.
On the remaining qubits $\neg S := \{1, ..., n\} \setminus S$, the gate acts trivially.
In particular, the distribution of outcomes $x_i$ for these trivial qubits must be the same for $P_k$ and $P_{k-1}$.
In other words, the marginal distribution
$
	P_k(\vec x_{\neg S}) := \sum_{\vec y_S} P_k(\vec x_{\neg S}, \vec y_S)
$
including only the qubits in $\neg S$ satisfies $P_k(\vec x_{\neg S}) = P_{k-1}(\vec x_{\neg S})$.
Therefore,
\[
	P_k(\vec{x})
	= P_k(\vec x_S ~|~ \vec x_{\neg S}) P_k(\vec x_{\neg S})
	= P_k(\vec x_S ~|~ \vec x_{\neg S}) P_{k-1}(\vec x_{\neg S}).
\]
Thus, in order to turn a sample $\vec y \sim P_{k-1}$ into a sample $\pvec y' \sim P_k$, the trivial bits can be kept unchanged, $\pvec y'_{\neg S} := \vec y_{\neg S}$, and only the bits $\pvec y'_S$ that $U_k$ acts on need to be resampled according to the conditional distribution
\[
	P_k(\pvec y'_S ~|~ \vec y_{\neg S})
	= \frac{P_k(\pvec y'_S, \vec y_{\neg S})}{\sum_{\vec x_S} P_k(\vec x_S, \vec y_{\neg S})}.
\]
For example, if $U_k$ is a single-qubit gate, then $S = \{ q \}$ for some qubit $q$.
To resample $y'_q$, one only needs to compute the two probabilities $p_0 = P_k(x_q=0, \vec x_{\neg q} = \vec y_{\neg q})$ and $p_1 = P_k(x_q=1, \vec x_{\neg q} = \vec y_{\neg q})$, and then set $y'_q := 1$ with probability $p_1 / (p_0 + p_1)$.
In general, updating a sample after applying an $\ell$-qubit gate requires $2^\ell$ amplitude evaluations, making the gate-by-gate algorithm scale exponentially in the maximum gate size.
However, in many cases this size is bounded, for example when considering circuits made up of standard single- and two-qubit gates.

Furthermore, there is a special case where amplitude evaluations can be skipped completely:
Suppose $U_k$ is a \emph{monomial gate}, meaning it maps computational basis states to other computational basis states up to a phase.\footnote{We call these gates monomial since they correspond to monomial matrices, i.e. matrices where each row and column has exactly one non-zero entry.}
In other words, suppose there are functions $f : \mathbb F_2^n \to \mathbb F_2^n$ and $\varphi : \mathbb F_2^n \to \mathbb R$ such that $U_k\ket{\vec x} = e^{i\varphi(\vec x)}\ket{f(\vec x)}$ for all $\vec x$.
Examples of these monomial gates include CNOT, Toffoli, T gates and (controlled) $Z$ rotations.
Note that $f$ must be injective since $U_k$ is unitary with $U_k^\dagger \ket{\vec x} = e^{-i\varphi(\vec x)}\ket{f^{-1}(\vec x)}$ such that
\[
	P_{k-1}(\vec x)
	= \left| \bra{\vec x} C_{k-1} \ket{0} \right|^2
	= \left| \bra{\vec x} U_k^\dagger U_k C_{k-1} \ket{0} \right|^2
	= \left| e^{-i\varphi(\vec x)} \bra{f^{-1}(\vec x)} U_kC_{k-1} \ket{0} \right|^2
	= P_k(f^{-1}(\vec x)).
\]
Thus, given a sample $\vec y \sim P_{k-1}$, it can be deterministically updated to $\pvec y' := f(\vec y)$, which satisfies $P_k(\pvec y') = P_{k-1}(\vec y)$ without needing to perform amplitude calculations.
Therefore, expensive resampling is only required when we encounter non-monomial gates such as Hadamards.
In particular, if the circuit is given in the CNOT+Hadamard+T gate set, then gate-by-gate sampling requires exactly $2h$ amplitude evaluations, where $h$ is the number of Hadamards.
The goal of the T-by-T sampling procedure that will be introduced in the next section is to remove this dependence on $h$ and instead make it scale only with the T-count $t$.

\section{T-by-T Sampling} \label{sec:t-by-t}

At a high level, T-by-T sampling follows a structure similar to the gate-by-gate method.
However, instead of processing the circuit one general gate at a time, it is sampled one \emph{T gate} at the time.
Concretely, the idea is to start with a pure Clifford circuit that does not contain any T gates.
Obtaining a sample from this initial circuit can be done efficiently using standard methods~\cite{aaronson2004improved,gidney2021stim}.
This sample is then refined by reintroducing the $T$ gates into the computation one by one, until we obtain a sample from the original circuit.
In \Cref{sec:algorithm} below, we give a description of the algorithm using a framework similar to \Cref{sec:gate-by-gate}.
The complexity of the algorithm is analysed in \Cref{sec:analysis}
Finally, \Cref{sec:mqbc} gives an alternative, more circuit-centric interpretation of T-by-T sampling and points out some connections to measurement based quantum computation.

\subsection{The Algorithm} \label{sec:algorithm}

Let $C$ be a Clifford+T circuit with T-count $t$.
Similar to gate-by-gate sampling, the algorithm proceeds by considering a sequence $\{C_k\}$ of circuits derived from $C$.
In particular, let $C_k$ be a modified version of $C$ that keeps only the first $k$ T gates and replaces the remaining $t-k$ ones with Z basis measurements.
In particular, $C_0$ is a Clifford computation and $C_t = C$.
Sampling from $C_k$ yields the final outcomes $\vec x \in \mathbb F_2^n$, but also $t-k$ additional bits $\vec a \in \mathbb F_2^{t-k}$ for the intermediate measurement results.
We write $C_k^{\vec a}$ for the circuit where these intermediate measurements are replaced with projectors $\Pi^{a_i} := \ket{a_i}\bra{a_i}$ onto the outcomes $\vec a$.
This yields a distribution $P_k(\vec x, \vec a) = \left| \bra{\vec x} C_k^{\vec a} \ket{0} \right|^2$ over the outcomes $\vec x, \vec a$.
The idea behind T-by-T sampling is to start with a sample from $P_0$ which can be computed efficiently.
Then, we refine this sample by inserting $T$ gates into the computation one by one until we arrive at a sample from the final distribution $P_t = P$.
Similar to gate-by-gate sampling, this works because adjacent distributions $P_{k-1}$ and $P_k$ are closely related.
Consider the following lemma:

\begin{lemma} \label{lem:flip}
	There exist bit strings $\vec\gamma_k \in \mathbb F_2^{n}$ and $\vec\delta_k \in \mathbb F_2^{t-k}$ such that for all $\vec x \in \mathbb F_2^{n}$ and $\vec a \in \mathbb F_2^{t-k}$
	\begin{equation} \label{eqn:flip-lem}
		P_{k-1} (\vec x, \vec a, a_k=0) + P_{k-1}(\vec x, \vec a, a_k=1)
		= \frac{1}{2} \left( P_k(\vec x, \vec a) + P_k(\vec x \oplus \vec\gamma_k, \vec a \oplus \vec\delta_k) \right).
	\end{equation}
\end{lemma}
\begin{proof}
	By definition
	\[
		C_{k-1}^{\vec a, a_k} = V^{\vec a} \Pi^{a_k}_q U
		\qquad\qquad
		C_k^{\vec a} = V^{\vec a} T_q U
	\]
	where $U$ is the Clifford+T prefix of $C_{k-1}^{\vec a, a_k}$ containing the first $k-1$ T gates, $\Pi^{a_k}$ is the projector replacing the $k$-th T gate on qubit $q$, and $V^{\vec a}$ is the remaining Clifford suffix of $C_{k-1}^{\vec a, a_k}$ with projectors replacing the remaining $t-k$ T gates according to outcomes $\vec a$.
	Using the fact that
	\begin{equation} \label{eqn:projector}
		\Pi^{a_k} = \ket{a_k}\bra{a_k} = \frac{1}{2}e^{-ia_k\frac{\pi}{4}}(T + (-1)^{a_k}ZT),
	\end{equation}
	we get
	\begin{align*}
		\sum_{a_k} P_{k-1}(\vec x, \vec a, a_k)
		&= \sum_{a_k} \left| \bra{\vec x} V^{\vec a} \Pi^{a_k}_q U \ket{0} \right|^2 \\
		&\overset{(\ref{eqn:projector})}{=} \sum_{a_k} \left| \frac{1}{2}e^{-ia_k\frac{\pi}{4}} \bra{\vec x} V^{\vec a} (T + (-1)^{a_k}ZT)_q U \ket{0} \right|^2 \\
		&= \sum_{a_k} \frac{1}{4} \left| \bra{\vec x} V^{\vec a} (T + (-1)^{a_k}ZT)_q U \ket{0} \right|^2 \\
		&= \sum_{a_k} \frac{1}{4} \left(
			\bra{\vec x}V^{\vec a}T_qU\ket{0} \bra{0}U^\dagger T_q^\dagger V^{\vec a \dagger}\ket{\vec x}
			+ (-1)^{a_k} \bra{\vec x}V^{\vec a}T_qU\ket{0} \bra{0}U^\dagger T_q^\dagger Z_q V^{\vec a \dagger}\ket{\vec x} \right. \\
			&\phantom{= \sum_{a_k} \frac{1}{4} ~~} \left.
			+ (-1)^{a_k} \bra{\vec x}V^{\vec a}Z_qT_qU\ket{0} \bra{0}U^\dagger T_q^\dagger V^{\vec a \dagger}\ket{\vec x}
			+ \bra{\vec x}V^{\vec a}Z_qT_qU\ket{0} \bra{0}U^\dagger T_q^\dagger Z_q V^{\vec a \dagger}\ket{\vec x}
		\right) \\
		&= \frac{1}{2} \bra{\vec x}V^{\vec a}T_qU\ket{0} \bra{0}U^\dagger T_q^\dagger V^{\vec a \dagger}\ket{\vec x}
			+ \frac{1}{2} \bra{\vec x}V^{\vec a}Z_qT_qU\ket{0} \bra{0}U^\dagger T_q^\dagger Z_q V^{\vec a \dagger}\ket{\vec x} \\
		&= \frac{1}{2} \left| \bra{\vec x} V^{\vec a} T_q U \ket{0} \right|^2
			+ \frac{1}{2} \left| \bra{\vec x} V^{\vec a} Z_qT_q U \ket{0} \right|^2
	\end{align*}
	Note that while the first term already corresponds to $\frac{1}{2}P_k(\vec x, \vec y)$, the second term has the wrong form since $Z_qT_q$ is applied instead of the desired $T_q$.
	However, since $V^{\vec a}$ is Clifford, the Pauli $Z$ can be pushed through it, potentially flipping some of the projector outcomes.
	In particular, there is a Pauli string $P$ and flip vector $\vec\delta_k \in \mathbb F_2^{t-k}$ such that $V^{\vec a}Z_q = PV^{\vec a \oplus \vec\delta_k}$ up to a phase.
	Similarly, there is a flip vector $\vec\gamma_k \in \mathbb F_2^{n}$ such that $\bra{\vec x}P = \bra{\vec x \oplus \vec\gamma_k}$ up to a phase.
	Therefore,
	\[
		\left| \bra{\vec x} V^{\vec a} Z_qT_q U \ket{0} \right|^2 
		= \left| \bra{\vec x \oplus \vec\gamma_k} V^{\vec a \oplus \vec\delta_k} T_q U \ket{0} \right|^2
		= P_k(\vec x \oplus \vec\gamma_k, \vec a \oplus \vec\delta_k).
	\]
\end{proof}

\Cref{lem:flip} tells us that the marginal distribution $P_{k-1}(\vec x, \vec a)$ over $a_k$ is an equal mixture of two copies of $P_k$, where one of them is shifted by the bit-flip vectors $\vec\gamma_k$ and $\vec\delta_k$.
Therefore, to refine a sample $(\vec y, \vec b, b_k) \sim P_{k-1}$ to the new distribution $P_k$, it suffices to choose between these two copies with the appropriate relative weight.
To this end, let
\begin{equation} \label{eqn:flip-prob}
	p(\vec y, \vec b) := \frac{P_k(\vec y \oplus \vec\gamma_k, \vec b \oplus \vec\delta_k)}{P_k(\vec y, \vec b) + P_k(\vec y \oplus \vec\gamma_k, \vec b \oplus \vec\delta_k)}
\end{equation}
be the flip probability for the sample $(\vec y, \vec b)$.
Similar to the gate-by-gate algorithm, $p(\vec y, \vec b)$ can be evaluated by computing the two probabilities $P_k(\vec y, \vec b)$ and $P_k(\vec y \oplus \vec\gamma_k, \vec b \oplus \vec\delta_k)$.
Then, with probability $p(\vec y, \vec b)$, set $\pvec y' := \vec y \oplus \vec\gamma_k$ and $\pvec b' := \vec b \oplus \vec\delta_k$.
Otherwise, keep $\pvec y' := \vec y$ and $\pvec b' := \vec b$.
The claim is that $(\pvec y', \pvec b')$ is now correctly distributed w.r.t. $P_k$.
To see this, let $Q(\pvec x', \pvec a')$ be the distribution of outcomes that we obtain using this process.
Note that there are exactly two ways we can end up with a given final configuration $(\pvec x', \pvec a')$:
\begin{enumerate}[{(i)}]
	\item either, we already sampled $(\pvec x', \pvec a')$ from $P_{k-1}$ and did not apply the flip, or
	\item we sampled $(\pvec x' \oplus \vec\gamma_k, \pvec a' \oplus \vec\delta_k)$ from $P_{k-1}$ and then applied the flip.
\end{enumerate}
Thus, we have
\begin{align*}
	Q(\pvec x', \pvec a')
	&= P_{k-1}(\vec x', \vec a')(1 - p(\vec x', \vec a')) + P_{k-1}(\pvec x' \oplus \vec\gamma_k, \pvec a' \oplus \vec\delta_k)p(\pvec x' \oplus \vec\gamma_k, \pvec a' \oplus \vec\delta_k) \\
	&= \frac{P_{k-1}(\vec x', \vec a') P_k(\pvec x', \pvec a')}{P_k(\pvec x', \pvec a') + P_k(\pvec x' \oplus \vec\gamma_k, \pvec a' \oplus \vec\delta_k)} + \frac{P_{k-1}(\pvec x' \oplus \vec\gamma_k, \pvec a' \oplus \vec\delta_k) P_k(\pvec x', \pvec a')}{P_k(\pvec x', \pvec a') + P_k(\vec x' \oplus \vec\gamma_k, \pvec a' \oplus \vec\delta_k)} \\
	&\overset{(\ref{eqn:flip-lem})}{=} \frac{P_{k-1}(\vec x', \vec a') P_k(\pvec x', \pvec a')}{2P_{k-1}(\vec x', \vec a')} + \frac{P_{k-1}(\pvec x' \oplus \vec\gamma_k, \pvec a' \oplus \vec\delta_k) P_k(\pvec x', \pvec a')}{2P_{k-1}(\pvec x' \oplus \vec\gamma_k, \pvec a' \oplus \vec\delta_k)} \\
	&= \frac{1}{2} P_k(\pvec x', \pvec a') + \frac{1}{2} P_k(\pvec x', \pvec a') \\
	&= P_k(\pvec x', \pvec a'),
\end{align*}
showing that $\pvec y', \pvec b'$ is indeed a correct sample from $P_k$.
To summarise, for each $T$ gate $k$ that is added back into the circuit, the current sample is randomly flipped by some vectors $\vec\gamma_k, \vec\delta_k$.
In particular, the flip probability $p$ at each step is computed from the relative weight between the probabilities $P_k(\vec x, \vec a)$ and $P_k(\vec x \oplus \vec\gamma_k, \vec a \oplus \vec\delta_k)$ as specified in equation (\ref{eqn:flip-prob}).
At the end, we arrive at a sample that is correctly distributed w.r.t. the target distribution $P_t = P$ of the original Clifford+T circuit.
Pseudocode for the entire sampling procedure is given in \Cref{alg:t-by-t}.

\begin{algorithm}[t]
	\caption{T-by-T Sampling}
	\label{alg:t-by-t}
	\begin{algorithmic}[1]
		\Statex \textbf{Input:} An $n$-qubit Clifford+$T$ circuit $C$ with $T$-count $t$
		\Statex \textbf{Output:} Sample $\vec x$ with probability $\left| \bra{\vec x}C\ket{0} \right|^2$
		
		\State $C_0 \gets$ Replace all $T$ gates in $C$ with $Z$ measurements
		\State $(\vec x, \vec a) \gets $ Sample from Clifford $C_0$
		
		\For{$k = 1$ \textbf{to} $t$}
		
		\State $(\vec a, a_k) \gets \textsc{Pop}(\vec a)$ \Comment{Drop outcome $a_k$}
		
		\State $C_k \gets$ Add next $T$ gate back into $C_{k-1}$
		\State $(\vec\gamma_k, \vec\delta_k) \gets$ Compute flip vectors according to \Cref{lem:flip}
		
		\State $A \gets \textsc{Prob}(C_k,\vec x,\vec a)$
		\Comment{Compute $P_k(\vec x, \vec a)$}
		\State $B \gets \textsc{Prob}(C_k,\vec x \oplus \vec\gamma_k, \vec a \oplus \vec\delta_k)$
		\Comment{Compute $P_k(\vec x \oplus \vec\gamma_k, \vec a \oplus \vec\delta_k)$}
		\State $p \gets B / (A + B)$
		
		\If{$\textsc{RandomSample}(p)$}
		\State $(\vec x, \vec a) \gets (\vec x \oplus \vec\gamma_k, \vec a \oplus \vec\delta_k)$
		\Comment{Apply flip with probability $p$}
		\EndIf
		
		\EndFor
		
		\State \Return $\vec x$
	\end{algorithmic}
\end{algorithm}

\subsection{Analysis} \label{sec:analysis}

The time complexity of T-by-T sampling heavily depends on the method used for evaluating the probabilities $P_k(\vec x, \vec a)$.
In general, for a circuit with T-count $t$, the algorithm will make exactly $2t$ calls to this strong simulation procedure.
Gate-by-gate sampling on the other hand requires two amplitude evaluations per non-monomial gate in the circuit.
Assuming that we are working with polynomially sized circuits, this corresponds to a polynomial number of strong simulations.
For simulation via low rank stabiliser decompositions, computing an amplitude for a circuit with T-count $t$ requires $2^{\alpha t}$ stabiliser terms.
In the case of gate-by-gate sampling, the worst case would be a circuit where the non-monomial gates occur after all $T$ have already been applied, making every evaluation cost the maximum of $2^{\alpha t}$ stabiliser terms, resulting in $O(\text{poly}(n)2^{\alpha t})$ terms in total.
For T-by-T sampling, the T-count increases gradually in each step, requiring only $\sum_{k=1}^t 2\cdot 2^{\alpha k} = O(2^{\alpha t})$ terms in total, thus yielding a polynomial advantage over gate-by-gate sampling.
Of course, the asymptotic growth is still dominated by the exponential scaling in $t$, in particular for the final evaluation of the circuit containing all T gates.
However, in situations where the input circuit contains a lot more non-monomial gates than T gates, T-by-T sampling requires significantly fewer calls to this exponentially costly procedure than gate-by-gate sampling, which makes it a lot faster in practice.

\subsection{MBQC Perspective} \label{sec:mqbc}

Finally, we give an alternative perspective on T-by-T sampling using ideas from \emph{measurement based quantum computation} (MBQC)~\cite{raussendorf2001oneway} and further relate it to gate-by-gate sampling.
First, note that Clifford+T circuits can be viewed as pure Clifford computations that consume $\ket{T} := \frac{1}{\sqrt 2}(\ket{0} + e^{i\frac{\pi}{4}}\ket{1})$ magic states.
In particular, those magic states can be turned back into $T$ gates using the following injection circuit:
\begin{equation} \label{eqn:t-inject}
	\begin{tikzpicture}
	\begin{pgfonlayer}{nodelayer}
		\node [style=none] (0) at (-1, 0.75) {};
		\node [style=none] (2) at (-1, -1) {};
		\node [style=none] (3) at (3.25, -1) {};
		\node [style=cnot ctrl] (4) at (0, -1) {};
		\node [style=cnot targ] (5) at (0, 0.75) {};
		\node [style=gate] (7) at (2, 0.75) {Meas};
		\node [style=gate] (8) at (2, -1) {S};
		\node [style=none] (13) at (-1.75, 0.75) {$\ket{T}$};
	\end{pgfonlayer}
	\begin{pgfonlayer}{edgelayer}
		\draw (3.center) to (2.center);
		\draw (5) to (4);
		\draw (7) to (0.center);
		\draw [style=double edge] (8) to (7);
	\end{pgfonlayer}
\end{tikzpicture}
 ~=~ \begin{tikzpicture}
	\begin{pgfonlayer}{nodelayer}
		\node [style=none] (0) at (26, 0) {};
		\node [style=none] (1) at (28, 0) {};
		\node [style=gate] (2) at (27, 0) {$T$};
	\end{pgfonlayer}
	\begin{pgfonlayer}{edgelayer}
		\draw [in=360, out=180] (1.center) to (0.center);
	\end{pgfonlayer}
\end{tikzpicture}

\end{equation}
Instead of state injection in its standard form, we may however also consider the time-reversed version of this protocol where, instead of preparing the ancilla in $\ket{T}$ and measuring in the $\{\ket{0}, \ket{1}\}$ basis, we prepare the ancilla in $\ket{0}$ and measure in the $\{ \ket{T_{+}}, \ket{T_{-}} \}$ basis where $\ket{T_{\pm}} := \frac{1}{\sqrt 2}(\ket{0} \pm e^{i\frac{\pi}{4}}\ket{1})$.
Equivalently, this corresponds to first applying $HT$ and then measuring in the computational basis:
\begin{equation} \label{eqn:t-mbqc}
	\begin{tikzpicture}
	\begin{pgfonlayer}{nodelayer}
		\node [style=none] (0) at (-1, 0.75) {};
		\node [style=none] (2) at (-1, -1) {};
		\node [style=none] (3) at (6.75, -1) {};
		\node [style=cnot ctrl] (4) at (0, -1) {};
		\node [style=cnot targ] (5) at (0, 0.75) {};
		\node [style=gate] (7) at (5.5, 0.75) {Meas};
		\node [style=gate] (8) at (5.5, -1) {Z};
		\node [style=none] (13) at (-1.75, 0.75) {$\ket{0}$};
		\node [style=gate] (14) at (1.5, 0.75) {$T$};
		\node [style=gate] (15) at (3.25, 0.75) {$H$};
	\end{pgfonlayer}
	\begin{pgfonlayer}{edgelayer}
		\draw (3.center) to (2.center);
		\draw (5) to (4);
		\draw (7) to (0.center);
		\draw [style=double edge] (8) to (7);
	\end{pgfonlayer}
\end{tikzpicture}
 ~=~ \begin{tikzpicture}
	\begin{pgfonlayer}{nodelayer}
		\node [style=none] (0) at (26, 0) {};
		\node [style=none] (1) at (28, 0) {};
		\node [style=gate] (2) at (27, 0) {$T$};
	\end{pgfonlayer}
	\begin{pgfonlayer}{edgelayer}
		\draw [in=360, out=180] (1.center) to (0.center);
	\end{pgfonlayer}
\end{tikzpicture}

\end{equation}
Notably, the correction for the $-1$ outcome is now $Z$ as opposed to $S$ in (\ref{eqn:t-inject}).
As an example, apply this transformation to the following circuit with two $T$ gates separated by a Clifford $V$:
\[ \input{./figures/0.tikz} ~\overset{(\ref{eqn:t-mbqc})}{=}~ \begin{tikzpicture}
	\begin{pgfonlayer}{nodelayer}
		\node [style=none] (10) at (8, 0.25) {};
		\node [style=none] (11) at (1.5, 0.25) {};
		\node [style=none] (14) at (8, -2.75) {};
		\node [style=none] (15) at (8, 0.75) {};
		\node [style=none] (16) at (9.5, -2.75) {};
		\node [style=none] (17) at (9.5, 0.75) {};
		\node [style=none] (18) at (8.75, -1) {$V$};
		\node [style=none] (19) at (1.5, -1) {};
		\node [style=none] (20) at (8, -1) {};
		\node [style=none] (22) at (1.5, -2.25) {};
		\node [style=none] (23) at (8, -2.25) {};
		\node [style=none] (31) at (16, 0.25) {};
		\node [style=none] (32) at (9.5, 0.25) {};
		\node [style=none] (40) at (9.5, -1) {};
		\node [style=none] (41) at (16, -1) {};
		\node [style=none] (43) at (9.5, -2.25) {};
		\node [style=none] (44) at (16, -2.25) {};
		\node [style=none] (53) at (1.5, 2.75) {};
		\node [style=cnot ctrl] (56) at (2.25, -1) {};
		\node [style=cnot targ] (57) at (2.25, 2.75) {};
		\node [style=gate] (58) at (6.75, 2.75) {Meas};
		\node [style=gate] (59) at (6.75, -1) {Z};
		\node [style=none] (60) at (0.75, 2.75) {$\ket{0}$};
		\node [style=gate] (61) at (3.25, 2.75) {$T$};
		\node [style=gate] (62) at (4.75, 2.75) {$H$};
		\node [style=none] (63) at (1.5, 1.5) {};
		\node [style=cnot ctrl] (64) at (10.25, -2.25) {};
		\node [style=cnot targ] (65) at (10.25, 1.5) {};
		\node [style=gate] (66) at (14.75, 1.5) {Meas};
		\node [style=gate] (67) at (14.75, -2.25) {Z};
		\node [style=none] (68) at (0.75, 1.5) {$\ket{0}$};
		\node [style=gate] (69) at (11.25, 1.5) {$T$};
		\node [style=gate] (70) at (12.75, 1.5) {$H$};
	\end{pgfonlayer}
	\begin{pgfonlayer}{edgelayer}
		\draw (11.center) to (10.center);
		\draw (15.center) to (14.center);
		\draw (17.center) to (15.center);
		\draw (14.center) to (16.center);
		\draw (16.center) to (17.center);
		\draw (20.center) to (19.center);
		\draw (23.center) to (22.center);
		\draw (32.center) to (31.center);
		\draw (41.center) to (40.center);
		\draw (44.center) to (43.center);
		\draw (57) to (56);
		\draw (58) to (53.center);
		\draw [style=double edge] (59) to (58);
		\draw (65) to (64);
		\draw (66) to (63.center);
		\draw [style=double edge] (67) to (66);
	\end{pgfonlayer}
\end{tikzpicture}
 \]
Since $V$ is Clifford, the first $Z$ correction can actually be delayed by pushing it through $V$ and the following CNOT, resulting in some Pauli string $P$ on the remaining qubits:
\[ \begin{tikzpicture}
	\begin{pgfonlayer}{nodelayer}
		\node [style=none] (10) at (3, 1.25) {};
		\node [style=none] (11) at (1, 1.25) {};
		\node [style=none] (14) at (3, -1.75) {};
		\node [style=none] (15) at (3, 1.75) {};
		\node [style=none] (16) at (4.5, -1.75) {};
		\node [style=none] (17) at (4.5, 1.75) {};
		\node [style=none] (18) at (3.75, 0) {$V$};
		\node [style=none] (19) at (1, 0) {};
		\node [style=none] (20) at (3, 0) {};
		\node [style=none] (22) at (1, -1.25) {};
		\node [style=none] (23) at (3, -1.25) {};
		\node [style=none] (31) at (17, 1.25) {};
		\node [style=none] (32) at (4.5, 1.25) {};
		\node [style=none] (40) at (4.5, 0) {};
		\node [style=none] (41) at (17, 0) {};
		\node [style=none] (43) at (4.5, -1.25) {};
		\node [style=none] (44) at (17, -1.25) {};
		\node [style=none] (53) at (1, 4) {};
		\node [style=cnot ctrl] (56) at (2.25, 0) {};
		\node [style=cnot targ] (57) at (2.25, 4) {};
		\node [style=gate] (58) at (10.625, 4) {Meas};
		\node [style=none] (60) at (0.25, 4) {$\ket{0}$};
		\node [style=gate] (61) at (7.25, 4) {$T$};
		\node [style=gate] (62) at (8.75, 4) {$H$};
		\node [style=none] (63) at (1, 2.5) {};
		\node [style=cnot ctrl] (64) at (5.25, -1.25) {};
		\node [style=cnot targ] (65) at (5.25, 2.5) {};
		\node [style=gate] (66) at (15.75, 2.5) {Meas};
		\node [style=gate] (67) at (15.75, -1.25) {Z};
		\node [style=none] (68) at (0.25, 2.5) {$\ket{0}$};
		\node [style=gate] (69) at (12.25, 2.5) {$T$};
		\node [style=gate] (70) at (13.75, 2.5) {$H$};
		\node [style=none] (71) at (10, 3) {};
		\node [style=none] (72) at (11.25, 3) {};
		\node [style=none] (73) at (10, -1.75) {};
		\node [style=none] (74) at (11.25, -1.75) {};
		\node [style=none] (75) at (10, 2.5) {};
		\node [style=none] (76) at (11.25, 2.5) {};
		\node [style=none] (77) at (10, 1.25) {};
		\node [style=none] (78) at (11.25, 1.25) {};
		\node [style=none] (79) at (10, 0) {};
		\node [style=none] (80) at (11.25, 0) {};
		\node [style=none] (81) at (10, -1.25) {};
		\node [style=none] (82) at (11.25, -1.25) {};
		\node [style=none] (83) at (10.625, 0.75) {$P$};
		\node [style=none] (84) at (10.625, 3.225) {};
		\node [style=none] (85) at (1.5, -2.25) {};
		\node [style=none] (86) at (6, -2.25) {};
		\node [style=none] (87) at (6, 4.75) {};
		\node [style=none] (88) at (1.5, 4.75) {};
	\end{pgfonlayer}
	\begin{pgfonlayer}{edgelayer}
		\draw (11.center) to (10.center);
		\draw (15.center) to (14.center);
		\draw (17.center) to (15.center);
		\draw (14.center) to (16.center);
		\draw (16.center) to (17.center);
		\draw (20.center) to (19.center);
		\draw (23.center) to (22.center);
		\draw (57) to (56);
		\draw (58) to (53.center);
		\draw (65) to (64);
		\draw [style=double edge] (67) to (66);
		\draw (73.center) to (71.center);
		\draw (74.center) to (73.center);
		\draw (74.center) to (72.center);
		\draw (72.center) to (71.center);
		\draw (75.center) to (63.center);
		\draw (76.center) to (66);
		\draw (77.center) to (32.center);
		\draw (78.center) to (31.center);
		\draw (79.center) to (40.center);
		\draw (80.center) to (41.center);
		\draw (81.center) to (43.center);
		\draw (82.center) to (44.center);
		\draw [style=double edge] (84.center) to (58);
		\draw [dashed] (87.center) to (86.center);
		\draw [dashed] (86.center) to (85.center);
		\draw [dashed] (88.center) to (85.center);
		\draw [dashed] (88.center) to (87.center);
	\end{pgfonlayer}
\end{tikzpicture}
 \]
Note that the beginning of the circuit in the dashed box is now purely Clifford and the T gates are applied at the very end, only interleaved with measurements and Pauli corrections.
By iterating this argument, it is easy to see that any Clifford+T circuit with T-count $t$ can be represented as some Clifford $W$ with $t$ additional ancillas initialised in $\ket{0}$, followed by $HT$ measurements and Pauli corrections:
\begin{equation} \label{eqn:mbqc-form}
	\begin{tikzpicture}
	\begin{pgfonlayer}{nodelayer}
		\node [style=none] (10) at (1, -2) {};
		\node [style=none] (11) at (0, -2) {};
		\node [style=none] (14) at (1, -4) {};
		\node [style=none] (15) at (1, 4) {};
		\node [style=none] (16) at (2.5, -4) {};
		\node [style=none] (17) at (2.5, 4) {};
		\node [style=none] (18) at (1.75, 0) {$W$};
		\node [style=none] (22) at (0, -3.5) {};
		\node [style=none] (23) at (1, -3.5) {};
		\node [style=none] (32) at (2.5, 0.25) {};
		\node [style=none] (40) at (2.5, -2) {};
		\node [style=none] (43) at (2.5, -3.5) {};
		\node [style=gate] (58) at (6.875, 3.5) {Meas};
		\node [style=gate] (61) at (3.5, 3.5) {$T$};
		\node [style=gate] (62) at (5, 3.5) {$H$};
		\node [style=none] (63) at (0, 2) {};
		\node [style=none] (68) at (-0.75, 2) {$\ket{0}$};
		\node [style=gate] (69) at (8.5, 2) {$T$};
		\node [style=gate] (70) at (10, 2) {$H$};
		\node [style=none] (71) at (6.25, 2.5) {};
		\node [style=none] (72) at (7.5, 2.5) {};
		\node [style=none] (73) at (6.25, -4) {};
		\node [style=none] (74) at (7.5, -4) {};
		\node [style=none] (75) at (6.25, 2) {};
		\node [style=none] (76) at (7.5, 2) {};
		\node [style=none] (77) at (6.25, 0.25) {};
		\node [style=none] (78) at (7.5, 0.25) {};
		\node [style=none] (79) at (6.25, -2) {};
		\node [style=none] (80) at (7.5, -2) {};
		\node [style=none] (81) at (6.25, -3.5) {};
		\node [style=none] (82) at (7.5, -3.5) {};
		\node [style=none] (83) at (6.875, -0.75) {$P_1$};
		\node [style=none] (84) at (6.875, 2.725) {};
		\node [style=none] (85) at (0.5, -2.5) {$\vdots$};
		\node [style=none] (86) at (1, 2) {};
		\node [style=none] (87) at (0, 3.5) {};
		\node [style=none] (88) at (-0.75, 3.5) {$\ket{0}$};
		\node [style=none] (89) at (1, 3.5) {};
		\node [style=none] (90) at (2.5, 3.5) {};
		\node [style=none] (91) at (2.5, 2) {};
		\node [style=none] (92) at (0, 0.25) {};
		\node [style=none] (93) at (-0.75, 0.25) {$\ket{0}$};
		\node [style=none] (94) at (1, 0.25) {};
		\node [style=gate] (95) at (11.875, 2) {Meas};
		\node [style=none] (96) at (11.25, 1) {};
		\node [style=none] (97) at (12.5, 1) {};
		\node [style=none] (98) at (11.25, -4) {};
		\node [style=none] (99) at (12.5, -4) {};
		\node [style=none] (102) at (11.25, 0.25) {};
		\node [style=none] (104) at (11.25, -2) {};
		\node [style=none] (106) at (11.25, -3.5) {};
		\node [style=none] (108) at (11.875, -1.25) {$P_2$};
		\node [style=none] (109) at (11.875, 1.225) {};
		\node [style=none] (111) at (12.5, -2) {};
		\node [style=none] (112) at (12.5, -3.5) {};
		\node [style=none] (114) at (16.25, -2) {};
		\node [style=none] (115) at (16.25, -3.5) {};
		\node [style=gate] (116) at (13.5, 0.25) {$T$};
		\node [style=gate] (117) at (15, 0.25) {$H$};
		\node [style=none] (118) at (12.5, 0.25) {};
		\node [style=gate] (119) at (16.875, 0.25) {Meas};
		\node [style=none] (120) at (16.25, -1.5) {};
		\node [style=none] (121) at (17.5, -1.5) {};
		\node [style=none] (122) at (16.25, -4) {};
		\node [style=none] (123) at (17.5, -4) {};
		\node [style=none] (124) at (16.25, -2) {};
		\node [style=none] (127) at (16.875, -2.75) {$P_t$};
		\node [style=none] (128) at (16.875, -1.275) {};
		\node [style=none] (129) at (18.5, -2) {};
		\node [style=none] (130) at (17.5, -2) {};
		\node [style=none] (131) at (17.5, -3.5) {};
		\node [style=none] (132) at (18.5, -3.5) {};
		\node [style=none] (133) at (18, -2.5) {$\vdots$};
		\node [style=none] (134) at (0.5, 1.25) {$\vdots$};
		\node [style=none] (135) at (14.5, -1) {$\ddots$};
		\node [style=none] (136) at (4.25, 1.25) {$\vdots$};
		\node [style=none] (137) at (4.25, -2.5) {$\vdots$};
		\node [style=none] (138) at (9.25, -2.5) {$\vdots$};
		\node [style=none] (139) at (14.5, -2.5) {$\vdots$};
	\end{pgfonlayer}
	\begin{pgfonlayer}{edgelayer}
		\draw (11.center) to (10.center);
		\draw (15.center) to (14.center);
		\draw (17.center) to (15.center);
		\draw (14.center) to (16.center);
		\draw (16.center) to (17.center);
		\draw (23.center) to (22.center);
		\draw (73.center) to (71.center);
		\draw (74.center) to (73.center);
		\draw (74.center) to (72.center);
		\draw (72.center) to (71.center);
		\draw (77.center) to (32.center);
		\draw (79.center) to (40.center);
		\draw (81.center) to (43.center);
		\draw [style=double edge] (84.center) to (58);
		\draw (63.center) to (86.center);
		\draw (87.center) to (89.center);
		\draw (90.center) to (58);
		\draw (75.center) to (91.center);
		\draw (92.center) to (94.center);
		\draw (98.center) to (96.center);
		\draw (99.center) to (98.center);
		\draw (99.center) to (97.center);
		\draw (97.center) to (96.center);
		\draw [style=double edge] (109.center) to (95);
		\draw (76.center) to (95);
		\draw (78.center) to (102.center);
		\draw (80.center) to (104.center);
		\draw (82.center) to (106.center);
		\draw (111.center) to (114.center);
		\draw (112.center) to (115.center);
		\draw (122.center) to (120.center);
		\draw (123.center) to (122.center);
		\draw (123.center) to (121.center);
		\draw (121.center) to (120.center);
		\draw [style=double edge] (128.center) to (119);
		\draw (118.center) to (119);
		\draw (130.center) to (129.center);
		\draw (132.center) to (131.center);
	\end{pgfonlayer}
\end{tikzpicture}

\end{equation}
This transformation can be viewed as a limited form of MBQC, where Cliffords remain as regular gates in the circuit model, but all T gates are realised via ancilla measurements.
Interestingly, the MBQC computation in (\ref{eqn:mbqc-form}) is closely related to the steps we described in \Cref{sec:algorithm}.
First, note that a sample from the initial Clifford $W$ exactly corresponds to a sample from the initial distribution $P_0(\vec x, \vec a)$:
All $T$ gates are replaced with CNOTs according to (\ref{eqn:t-mbqc}) with the target initialised in $\ket{0}$.
Thus, measuring one of the ancillas in the Z basis corresponds to applying a projector $\Pi^{a_k}$ instead of the $T$ gate and yields the additional outcome $a_k$.
After applying the change of basis $HT$, the measurement injects either a $T$ or a $ZT$ gate into the circuit, which is similar to equation (\ref{eqn:projector}) in the proof of \Cref{lem:flip}.
Finally, the X components of the Pauli corrections $P_k$ exactly describe the bit flips $\vec\gamma_k, \vec\delta_k$ from \Cref{lem:flip}.

From this perspective, once we have sampled from the initial Clifford circuit $W$, the remaining computation in (\ref{eqn:mbqc-form}) can be interpreted using the gate-by-gate framework.
The injected T gate itself is monomial and therefore preserves computational basis states. The subsequent Hadamard, however, is non-monomial and requires resampling.
Resampling after the Hadamard amounts to evaluating two amplitudes corresponding to projections onto $\bra{T_{+}}$ and $\bra{T_{-}}$.
These amplitudes coincide with those of the circuit in which either a $T$ or a $ZT$ gate is injected.
After commuting the $Z$ through the Clifford suffix, these are precisely the amplitudes $P_k(\vec x, \vec a)$ and $P_k(\vec x \oplus\vec\gamma_k, \vec a \oplus\vec\delta_k)$ from \Cref{sec:algorithm}.
Thus, \Cref{lem:flip} tells us that $P_{k-1}$ can be interpreted as the distribution coming from the equal mixture of the branches where either $T$ or $ZT$ is injected.
Then, the flip probability (\ref{eqn:flip-prob}) is exactly the posterior probability that a given sample arose from the $ZT$ branch rather than the $T$ branch.
If the $ZT$ branch is selected, we apply the deterministic bit flip that maps the sample back into the $T$ branch.
Thus, we can view the T-by-T update rule as projecting the ancilla measurements in (\ref{eqn:mbqc-form}) onto the +1 outcome.

Interestingly, the circuit (\ref{eqn:mbqc-form}) also shows us a slight alternative implementation of T-by-T sampling.
Instead of projecting onto the +1 outcome, we could keep the -1 outcome but then modify the circuit by adding the Pauli correction $P_k$.
Since Paulis are monomial gates, they just correspond to deterministic flips in the outcome and do not require resampling.
The main difference is that these Pauli corrections would remain in the circuit when we move on to the next T gates.
In particular, $X$ components in the correction frame would correspond to flipping future $T$ gates into $T^\dagger$ gates.
This complication is avoided by following the postselected algorithm as described in \Cref{sec:algorithm}.

\section{Experiments} \label{sec:experiments}

To compare T-by-T sampling against the gate-by-gate and qubit-by-qubit marginalisation method, we ran numerical experiments measuring the sampling time for both random circuits (\Cref{sec:random}) and structured QEC circuits for magic state cultivation (\Cref{sec:cultivation}).
Since the goal of this section is to compare sampling strategies rather than the underlying strong simulation techniques, the stabiliser decomposition backend is kept identical across all experiments.
Any observed performance differences therefore arise from the sampling strategy, specifically, from how many amplitude evaluations are required and at which effective T-count they are performed.
In particular, all experiments use the BSS + cat state stabiliser decomposition algorithm~\cite{kissinger2022classical} implemented in the \texttt{quizx} library~\cite{kissinger2022simulating, quizx}, achieving the state of the art worst-case scaling factor of $\alpha \approx 0.396$.
All tests were performed on a consumer laptop, imposing a 10 minute time limit per sample.
All data points represent running times on a single CPU core. 
Implementations of all sampling methods, together with code to generate the circuits and reproduce the results are open source and available at:
\begin{center}
	\url{https://github.com/mark-koch/t-by-t-sampling}
\end{center}

\subsection{Random Clifford+T Circuits} \label{sec:random}

First, we consider random Clifford+T circuits constructed from multiple layers of exponentiated Pauli unitaries of the form $e^{-i\frac{\pi}{8}P}$ for some dense Pauli string $P$ whose entries are uniformly sampled from $\{I,X,Y,Z\}$.
Each Pauli exponential contributes exactly one T gate to the circuit, so the T-count $t$ is equal to the number of layers.
For the qubit-by-qubit marginalisation method, this T-count will effectively double to $2t$ since computing marginals corresponds to evaluating amplitudes of a circuit with twice as many T gates.
For the gate-by-gate method, the T-count remains $t$, but each layer requires resampling of qubit outcomes $x_i$ for all positions where the Pauli support $P_i$ is $X$ or $Y$ as these require a non-monomial change of basis.
For dense Pauli strings, this occurs on approximately half of the qubits in expectation, so each layer requires $O(n)$ resampling steps, where $n$ is the number of qubits.
Over the entire circuit, this results in $O(nt)$ amplitude evaluations with the T-count increasing after each layer.
In contrast, T-by-T sampling only requires $2t$ amplitude evaluations in total, independent of $n$.

\begin{figure}
	\centering
	\begin{subfigure}[b]{0.49\textwidth}
		\centering
		\includegraphics[scale=0.7]{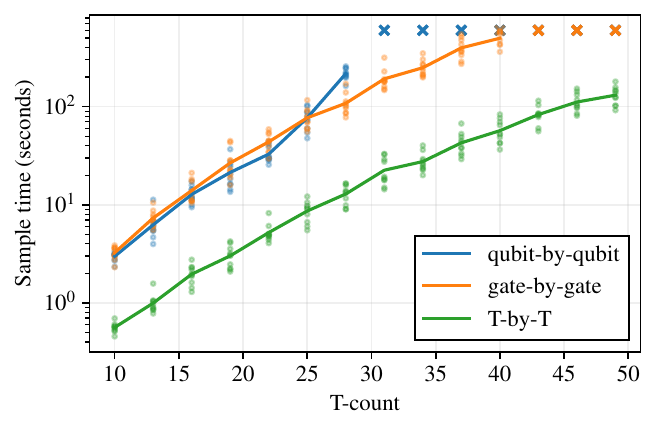}
		\caption{Random 50-qubit circuits with increasing T-counts} \label{fig:increasing-tcounts}
	\end{subfigure}
	\hfill
	\begin{subfigure}[b]{0.49\textwidth}
		\centering
		\includegraphics[scale=0.7]{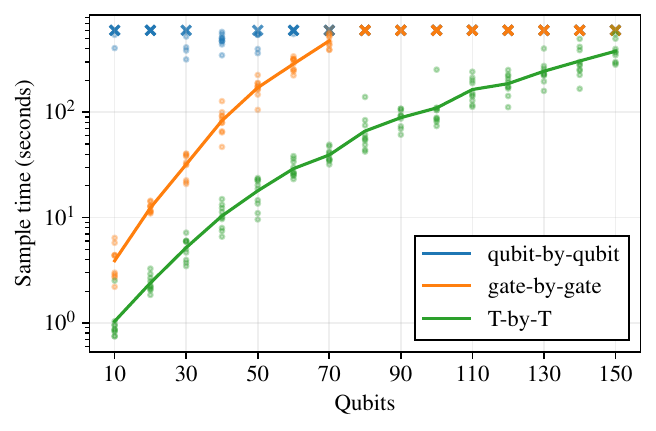}
		\caption{Random T-count 30 circuits with increasing qubits} \label{fig:increasing-qubits}
	\end{subfigure}
	\caption{Runtime for obtaining a sample from random Clifford+T circuits. The solid lines indicate an average over 10 runs. Cross marks denote runs that exceeded the 10 minute time limit.}
\end{figure}

\Cref{fig:increasing-tcounts} plots the runtime of these three methods for random 50-qubit circuits with increasing T-counts.
As discussed in \Cref{sec:analysis}, any sampling method based on stabiliser decompositions is expected to scale exponentially in $t$.
This behaviour is clearly visible in \Cref{fig:increasing-tcounts}: all three curves exhibit exponential growth in $t$.
Moreover, the gate-by-gate and T-by-T methods have essentially identical slopes.
This is because both methods perform amplitude evaluations at T-count $t$, which dominates the asymptotic scaling.
Nevertheless, T-by-T sampling consistently outperforms gate-by-gate sampling by roughly an order of magnitude across the tested range, since it only requires two evaluations at full T-count $t$.
Finally, qubit-by-qubit marginalisation remains competitive with gate-by-gate sampling only up to about $t \approx 25$, after which the exponential penalty from doubling the T-count becomes dominant.

In \Cref{fig:increasing-qubits}, the T-count is fixed to $t = 30$ and instead the number of qubits $n$ is varied, thereby also increasing the number of Clifford operations.
This isolates the scaling of each method in relation to the overall circuit size rather than the T-count.
In this setting, the advantage of T-by-T sampling becomes even clearer: the runtime gap between T-by-T and gate-by-gate sampling widens as the number of qubits increases, reflecting the 
$O(nt)$ resampling overhead of the gate-by-gate method.
Qubit-by-qubit marginalisation is no longer competitive in this setting, with most runs exceeding the allotted time limit.

\subsection{Magic State Cultivation Circuits} \label{sec:cultivation}

While random Clifford+T circuits are useful for studying generic scaling behaviour, they do not necessarily reflect the structure of real-world circuits encountered in practice.
To evaluate performance in a more realistic setting, we now turn to Clifford+T circuits from the recent quantum error correction (QEC) literature.
Concretely, we consider magic state cultivation circuits~\cite{gidney2024cultivation} implementing a recent protocol for constructing high fidelity $\ket{T}$ magic states within QEC codes.
These circuits consist of an initial injection and cultivation step that prepares and verifies a logical $\ket{T}$ state in a small low-distance code.
For the following experiments, we use the distance 3 injection and cultivation circuits from \cite{gidney2024cultivation}, which have a T-count of only $t = 15$.
Then, the circuit is composed with an escape stage that rapidly grows the code size.
Notably, the escape stage is purely Clifford, but it involves a large number of qubits, gates, and measurements.
Following \cite{gidney2024cultivation}, we consider escape circuits with three QEC rounds during the growth step and five rounds in the final code.
Overall circuit statistics for increasing escape distances are summarised in \Cref{tab:cultiv}.

\begin{figure}
	\centering
	\begin{subfigure}[b]{0.49\textwidth}
		\centering
		\includegraphics[scale=0.7]{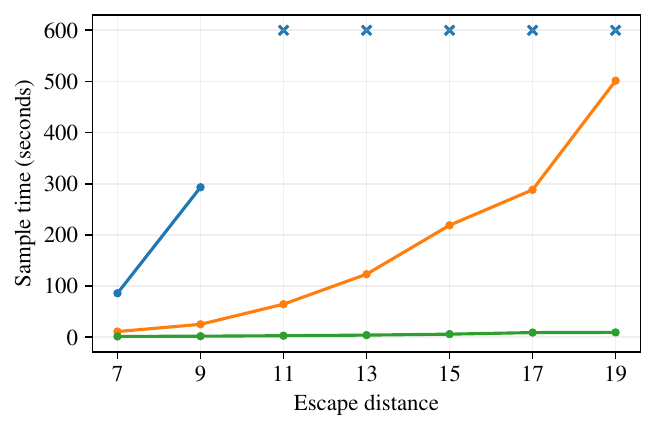}
	\end{subfigure}
	\hfill
	\begin{subfigure}[b]{0.49\textwidth}
		\centering
		\includegraphics[scale=0.7]{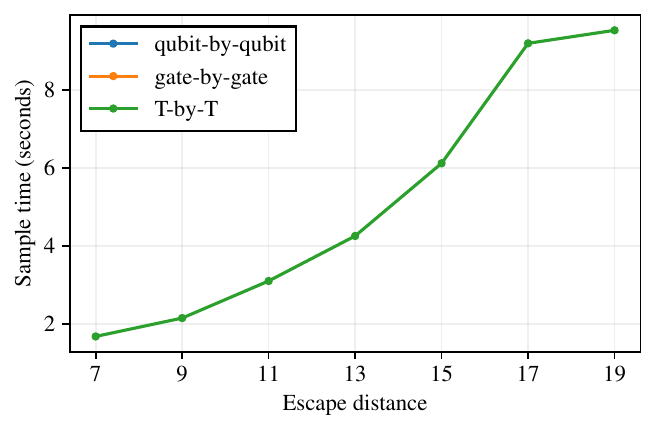}
	\end{subfigure}
	\caption{Runtime for obtaining a (noiseless) sample from distance 3 magic state cultivation circuits including escape into codes of increasing distance. The left side shows a zoomed in view of the scaling of T-by-T sampling. Cross marks denote runs that exceeded the 10 minute time limit.}
	\label{fig:cultiv}
\end{figure}

\begin{table}
	\centering
	\begin{tabular}{r|r|r|r|r|r}
		\toprule   
		\textbf{Distance} & \textbf{Qubits} & \textbf{Gates} & \textbf{Measurements} & \textbf{H Gates} & \textbf{T Gates} \\
		\midrule
		7  & 102 & 2,290  & 520  & 520  & 15 \\
		9  & 166 & 3,675  & 840  & 825  & 15 \\
		11 & 246 & 5,424  & 1,240 & 1,206 & 15 \\
		13 & 342 & 7,573  & 1,720 & 1,663 & 15 \\
		15 & 454 & 10,014 & 2,280 & 2,196 & 15 \\
		17 & 582 & 12,855 & 2,920 & 2,805 & 15 \\
		19 & 726 & 16,060 & 3,640 & 3,490 & 15 \\
		\bottomrule
	\end{tabular}
	\caption{Magic state cultivation circuit sizes for different escape distances.} \label{tab:cultiv}
\end{table}

\Cref{fig:cultiv} shows the sampling time for noiseless cultivation circuits with increasing escape distances.
Because the escape stage contains many qubits and a large number of Hadamard gates, both qubit-by-qubit marginalisation and gate-by-gate sampling incur substantial runtime overheads as the distance increases. In contrast, the T-by-T method requires only $2t = 30$ amplitude evaluations, independent of escape distance and the size of the surrounding Clifford computation.
As a result, the sampling time increases only moderately, from approximately 1.5s to 10s, as the escape distance grows from 7 to 19.
Gate-by-gate sampling, on the other hand, suffers an increase from roughly 11s to 500s over the same range.
Finally, the qubit-by-qubit method already times out after distance 9.
These results highlight that T-by-T sampling is particularly well suited for large-scale QEC simulations, where a small number of non-Clifford resources are embedded in large Clifford computations, making it a promising tool for simulating fault-tolerant protocols.

\section{Discussion}

In this work, we have shown that exact weak simulation of Clifford+T circuits can be reduced to a number of strong simulations that depends only on the T-count, and is independent of the surrounding Clifford structure.
We demonstrated the advantage of this new T-by-T sampling approach over existing methods for a class of random circuits as well as magic state cultivation circuits~\cite{gidney2024cultivation}, where other baselines are outperformed by one to two orders of magnitude.
However, it is important to emphasise that this advantage is not universal.
There exist classes of circuits for which alternative approaches may be more suitable.
For example, dense single-layer IQP circuits~\cite{bremner2016iqp} can contain significantly more T gates than Hadamard gates.
In such regimes, the gate-by-gate method may be preferable, as its cost scales with the number of Hadamard gates rather than the T-count.
T-by-T sampling, by contrast, is most effective in settings where a moderate number of T gates are embedded within large Clifford-dominated computations.

One such setting is simulation of large scale QEC experiments.
The magic state cultivation benchmarks in \Cref{sec:cultivation} are very promising and suggest that T-by-T sampling could become a valuable tool for this application.
In that context it is worth stressing that the simulations in \Cref{sec:cultivation} were single noiseless samples, whereas actual QEC simulations would typically insert stochastic noise across many shots, apply corrections conditioned on detected syndromes, and evaluate the overlap with the ideal logical state in order to determine the logical error rate.
However, the goal of this work was not to reproduce such an end-to-end QEC study, but rather to demonstrate that T-by-T sampling is particularly well suited to the circuit structures that arise in that context.

A natural next step for future work would therefore be to integrate T-by-T sampling into a complete QEC simulation pipeline.
Notably, there has been substantial recent progress in simulating Clifford+T QEC experiments, including techniques for pushing noise though circuits~\cite{surti2025efficient}, improved stabiliser decompositions for cultivation circuits~\cite{wan2026simulating}, and parametric decompositions that allow many shots to be sampled in parallel~\cite{tsim, sutcliffe2025parametric}.
Furthermore, Chase and Labib~\cite{chase2026clifft} propose a simulation approach where the state is factored into an offline stabiliser tableau and a small active state vector.
However, all of these works still rely on qubit-by-qubit marginalisation or similar approaches for sampling.
In that sense, the contribution of this work is largely orthogonal to these developments.
By combining T-by-T sampling with some of these other ideas, we can hopefully obtain further significant speedups for QEC simulations.

Finally, although this work focuses primarily on Clifford+T circuits, the T-by-T algorithm actually trivially generalises to arbitrary Clifford+$R_Z$ circuits.
However, in that broader setting, stabiliser decomposition techniques are typically only approximate rather than exact~\cite{bravyi2019simulation}.
Investigating the robustness of T-by-T sampling to approximation errors and applying it to Clifford+$R_Z$ circuits would therefore be another interesting direction for future work.

\enlargethispage{\baselineskip}
\paragraph{Acknowledgements}
We would like to thank Benjamin Rodatz for useful discussions that led to the discovery of this algorithm.
Furthermore, MK is supported by the Clarendon Fund Scholarship and the Sloane Robinson Scholarship.

\nocite{*}
\bibliographystyle{eptcs}
\bibliography{bibliography}
\end{document}